\documentclass[12pt,a4paper]{iopart}
\usepackage{iopams}             
\usepackage{graphicx}           
\usepackage{hyperref}           

\begin{document}

\title [Impact of the initial size of spatial fluctuations in Pb-Pb
  collisions at LHC] {Impact of the initial size of spatial
  fluctuations on the collective flow in Pb-Pb collisions at
  $\sqrt{s_{NN}}$ = 2.76 TeV}

\author{V P Konchakovski$^1$, W Cassing$^1$ and V D Toneev$^2$}
\address{$^1$ Institute for Theoretical Physics, University of Giessen, 35392 Giessen, Germany}
\address{$^2$ Joint Institute for Nuclear Research, 141980 Dubna, Russia}

\begin{abstract}
The Parton-Hadron-String-Dynamics (PHSD) transport model is used to
study the influence of the initial size of spatial fluctuations of the
interacting system on flow observables in Pb-Pb collisions at
$\sqrt{s_{NN}}$ = 2.76 TeV for different centralities. While the flow
coefficients $v_2$, $v_3$, $v_4$ and $v_5$ are reasonably described in
comparison to the data from the ALICE Collaboration for different
centralities within the default setting, no essential sensitivity is
found with respect to the initial size of spatial fluctuations even
for very central collisions where the flow coefficients are dominated
by the size of initial state fluctuations. We attribute this lack of
sensitivity partly to the low interaction rate of the
degrees-of-freedom in this very early phase of order $\sim$ 0.3 fm/c
which is also in common with the weakly interacting color glass
condensate (CGC) or glasma approach. Moreover, since the event shape
in the transverse plane is approximately the same for different size
of spatial fluctuations very similar eccentricities $\epsilon_n$ are
transformed to roughly the same flow coefficients $v_n$ in momentum
space.
\end{abstract}

\submitto{\JPG}
\maketitle

\section{Introduction}
\label{sec:intro}

Ultra-relativistic nucleus-nucleus collisions allow to study strongly
interacting QCD matter under extreme conditions in heavy-ion
experiments at the relativistic heavy-ion collider (RHIC) and the
large hadron collider (LHC). The experiments at the RHIC and the LHC
have demonstrated that a stage of partonic matter is produced in these
reactions which is in an approximate equilibrium for a couple of
fm/c~\cite{3years05,QM11}. Due to the non-perturbative and
non-equilibrium nature of relativistic nuclear reaction systems, their
theoretical description is based on a variety of effective approaches
ranging from hydrodynamic models with different initial
conditions~\cite{IdealHydro1, IdealHydro2, IdealHydro3, IdealHydro4,
  IdealHydro5, IdealHydro6, ViscousHydro1, ViscousHydro2,
  ViscousHydro3, ViscousHydro4} to various kinetic
approaches~\cite{T1, T2, T3, all, Greco, PhysRep, AMPT, BAMPS, CB09}
or different types of hybrid models~\cite{hyb1, hyb2, hyb3, hyb4,
  hyb5,hyb6, hyb7, hyb8}. In the latter hybrid approaches the initial
state models are followed by an ideal or viscous hydro phase which
after hadronic freeze-out is followed up by a hadronic transport
approach to take care of the final elastic and inelastic hadronic
reactions.

The actual question addressed in this study is whether the initial
size of spatial fluctuations of the colliding zone leaves its traces
in the collective flow coefficients $v_n$. A color glass condensate
(CGC)~\cite{GIJV10} is expected to lead to structures of smaller scale
or higher size of spatial fluctuations as compared to the Glauber
model that incorporates fluctuations on the nucleon scale. A similar
question has been addressed in the ideal or viscous hydro calculations
by the authors of Refs.~\cite{Schenke, Schenke0, Schenke12, Schenke13,
  Schenke14, Werner0, Werner} which have found some sensitivity with
respect to higher moments $v_n$.

In the ideal or viscous hydro calculations the initial conditions --
at some finite starting time of the order of 0.3 to 0.5 fm/c -- have
to be evaluated either in terms of the (standard) Glauber model or
other initial state scenarios like in the IP-glasma
model~\cite{Schenke,Schenke0} or the CGC approach,
respectively. Differences between the different initial state
assumptions and dynamical evolutions thus have to be expected. The
applicability of ideal or viscous hydrodynamic models to
proton-nucleus reactions for low multiplicity events, however, is very
much debated. This also holds for hybrid models as long as they employ
a hydro phase. To our knowledge only microscopic transport approaches
allow to bridge the gap from p-p to p-A and A-A collisions in a unique
way without introducing additional (and less controlled) parameters.

We recall that the flow harmonics $v_n$ for the azimuthal angular
distribution of hadrons have been found to be sensitive to the early
stage of the nuclear interaction and in particular to their
fluctuations. Indeed, the detailed heavy-ion analysis in
Ref.~\cite{Al11} shows that Monte Carlo CGC approaches (MC-CGC)
systematically give a larger initial eccentricity than Glauber
models. However, it is unclear to what extent such properties of the
CGC formalism are robust with respect to extended correlations. Also,
studies of higher harmonics -- as presented in~\cite{QM11} by the
PHENIX or ALICE collaborations -- do not clearly favor the CGC or
Glauber assumptions for the initial state of the collision. The first
LHC data on the bulk particle production in Pb-Pb collisions are in
good agreement with improved CGC expectations but they are also
compatible with Monte Carlo event
generators~\cite{3years05,ATLAS-fl13}.

The complexity of heavy-ion collisions is reduced essentially in the
case of proton-nucleus collisions due to the expected dominance of the
initial state effects over final state effects. Recently, we have
performed a microscopic transport study of p-Pb collisions at a
nucleon-nucleon center-of-mass energy $\sqrt{s_{NN}}=$ 5.02 TeV and
compared our results to the first preliminary ALICE measurement at the
LHC of the charged particle pseudorapidity distributions from
Ref.~\cite{ALICE:2012xs} for pseudorapidity $|\eta|<$~2 for different
multiplicity bins of charged particles
$N_{ch}$~\cite{Konchakovski:2014wqa}. However, these differential
pseudorapidity densities do not allow for firm conclusions on the
initial state configuration since independent approaches compare
reasonably well, too: the saturation models employing coherence
effects~\cite{DK12, TV12, Albacete:2012xq} or the two-component models
combining perturbative QCD processes with soft
interactions~\cite{BBG12, XDW12}. On the other side, a sizeable
difference in the mean transverse momentum of particles
$\left<p_T\right>$ versus the pseudorapidity $\eta$ with opposite
slopes in $\eta$ on the projectile side is found within the CGC
framework relative to hydrodynamical or transport
calculations~\cite{Konchakovski:2014wqa}.

We here explore the sensitivity of the collective flow coefficients
$v_n$ and related quantities on the initial size of spatial
fluctuations in Pb-Pb interactions at the collision energy
$\sqrt{s_{NN}}=$~2.76 TeV within the parton-hadron-string-dynamics
(PHSD) transport approach~\cite{CB09} which has been properly upgraded
to LHC energies with respect to a more recent PYTHIA 6.4
implementation~\cite{Konchakovski:2014wqa}. After a brief reminder of
the PHSD approach and generic results for transverse momentum spectra
and flow coefficients $v_n(p_T)$ for central and mid-central Pb-Pb
collisions in comparison to available data in Sec.~\ref{sec:phsd} we
present the actual results for $v_n(p_T)$ for very central Pb-Pb
collisions employing different sizes of spatial fluctuations in
Sec.~\ref{sec:granul}. We conclude our findings in
Sec.~\ref{sec:conclusions}.

\section{PHSD @ LHC}
\label{sec:phsd}

The PHSD model is a covariant dynamical approach for strongly
interacting systems formulated on the basis of Kadanoff-Baym
equations~\cite{JCG04} or off-shell transport equations in phase-space
representation, respectively. In the Kadanoff-Baym theory the field
quanta are described in terms of dressed propagators with complex
selfenergies. Whereas the real part of the selfenergies can be related
to mean-field potentials (of Lorentz scalar, vector or tensor type),
the imaginary parts provide information about the lifetime and/or
reaction rates of time-like particles~\cite{Ca09}. Once the proper
(complex) selfenergies of the degrees of freedom are known, the time
evolution of the system is fully governed by off-shell transport
equations (as described in Refs.~\cite{JCG04, Ca09}). This approach
allows for a simple and transparent interpretation of lattice QCD
results for thermodynamic quantities as well as correlators and leads
to effective strongly interacting partonic quasiparticles with broad
spectral functions. For a review on off-shell transport theory we
refer the reader to Ref.~\cite{Ca09}; model results and their
comparison with experimental observables for heavy-ion collisions from
the lower super-proton-synchrotron (SPS) to
relativistic-heavy-ion-collider (RHIC) energies can be found in
Refs.~\cite{CB09,To12,KB12,Kb12b} including electromagnetic probes
such as $e^+e^-$ or $\mu^+\mu^-$ pairs~\cite{el-m,LiLHC} or real
photons~\cite{photons}.

\subsection{Extensions @ LHC}

To extend the PHSD model to higher energies than $\sqrt{s_{NN}}=$
200~GeV at RHIC the PYTHIA 6.4 generator~\cite{Sjostrand:2006za} has
been additionally implemented for initial nucleon collisions at LHC
energies~\cite{Konchakovski:2014wqa}. For the subsequent (lower
energy) collisions the standard PHSD model~\cite{CB09} is applied
(including PYTHIA v5.5 with JETSET v7.3 for the production and
fragmentation of jets~\cite{PYTHIA0}, i.e.\ for $\sqrt{s_{NN}} \le $
500 GeV~\cite{PYTHIA0}). In this way all results from PHSD are
regained up to top RHIC energies and a proper extension to LHC
energies is achieved. At $\sim \sqrt{s_{NN}}=$ 500 GeV both PYTHIA
versions lead to very similar results. In PYTHIA 6.4 we use the
Innsbruck pp tune (390) which allows to describe reasonably the p-p
collisions at $\sqrt{s_{NN}}=$ 7 TeV in the framework of the PHSD
transport approach (cf.\ Fig.~1 in~\cite{Konchakovski:2014wqa}). The
overall agreement with LHC experimental data for the distribution in
the charged particle multiplicity $N_{ch}$, the charged particle
pseudorapidity distribution, the transverse momentum $p_T$ spectra and
the correlation of the average $p_T$ with the number of charged
particles $N_{ch}$ is satisfactory. Also a variety of observables from
p-Pb collisions at $\sqrt{s_{NN}}= $ 5.02 TeV compare quite well with
the experimental observations~\cite{Konchakovski:2014wqa}.

\begin{figure*}[th]
\centering
\includegraphics[width=0.7\textwidth]{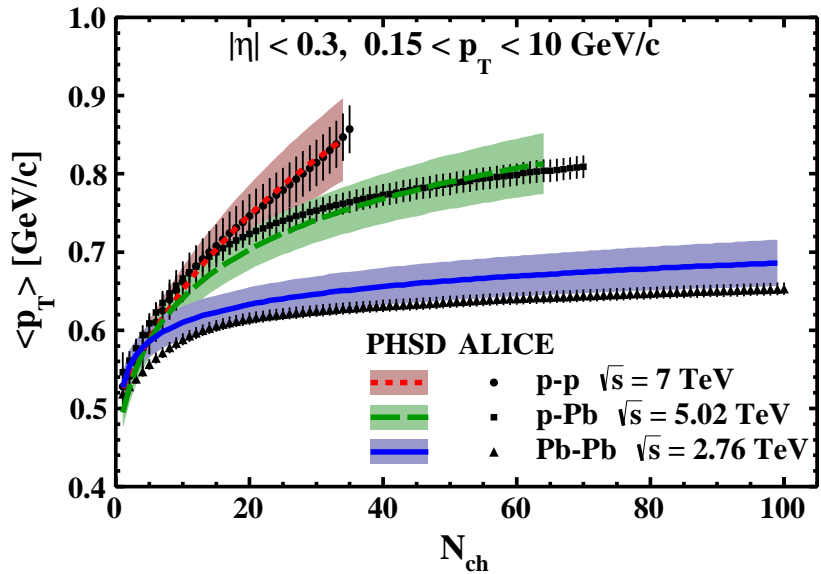}
\caption{Mean $p_T$ results for p-p, p-Pb and Pb-Pb collisions from
  the PHSD transport approach in comparison to the ALICE experimental
  data from Ref.~\cite{Abelev:2013bla}. Note the different invariant
  energies for p-p, p-Pb and Pb-Pb collisions.}
\label{fig:meanpt}
\end{figure*}

The first (homework) question to answer is whether the PHSD approach
still works at LHC energies for nucleus-nucleus collisions although
the invariant energy is higher by about a factor of 13.8 than at the
top RHIC energy. In Fig.~\ref{fig:meanpt} we compare the mean $p_T$ as
a function of charged multiplicity $N_{ch}$ in p-p reactions at
$\sqrt{s_{NN}}=$ 7 TeV, p-Pb collisions at $\sqrt{s_{NN}}=$ 5.02 TeV
and Pb-Pb collisions at $\sqrt{s_{NN}}=$ 2.76 TeV from the PHSD to the
experimental data from Ref.~\cite{Abelev:2013bla}. Note that for low
multiplicities ($N_{ch} < 5$) the mean $p_T$ is almost independent on
energy (see also Ref.~\cite{Abelev:2013bla}) which in PHSD can be
traced back to the fact that (for the acceptance $|\eta| \leq$ 0.3,
0.15 $\leq p_T \leq $ 10 GeV/c) only events with one or two binary
collisions $N_{bin}$ are selected for all systems.  Actually, the
correlation $<p_T>(N_{ch})$ only weakly depends on $\sqrt{s_{NN}}$ for
$pp$ reactions at these LHC energies, however, when plotting
$p_T(N_{ch})$ on an event-by-event basis, large fluctuations in $p_T$
or $N_{ch}$ are obtained within PHSD. The same holds true for p+Pb and
Pb-Pb reactions where a fixed $N_{ch}$ can be obtained by reactions
with a varying number of binary collisions $N_{bin}$. Each of these
binary reactions then has a low $N_{ch}$ and $<p_T>$,
respectively. The ensemble average finally leads to the average
correlation shown in Fig.~\ref{fig:meanpt}.  Nevertheless, the
agreement between data and calculations (within the statistical
accuracy) is encouraging. Note again that only very peripheral Pb-Pb
collisions are probed for $N_{ch} < $ 100.

\subsection{$p_T$-spectra for central Pb-Pb}
\label{sec:pt}

\begin{figure*}[th]
\centering
\includegraphics[width=0.7\textwidth]{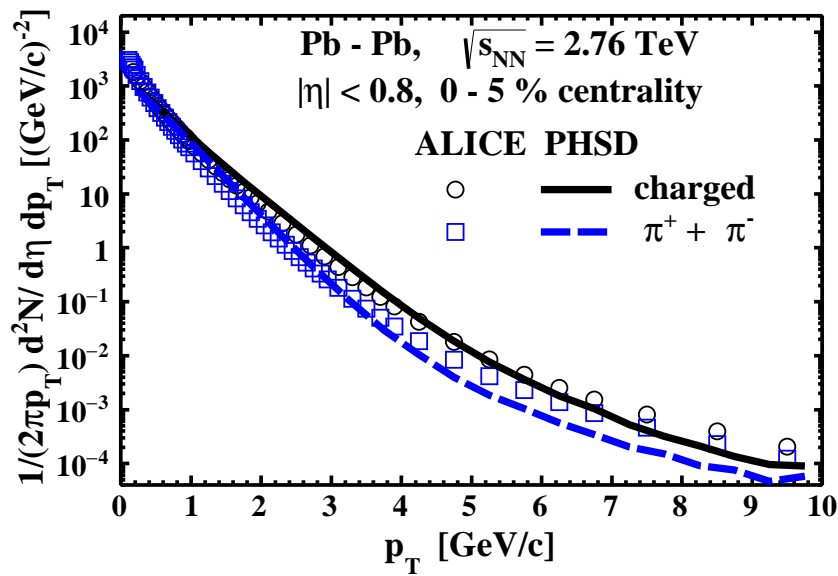}
\caption{Transverse momentum spectra from PHSD in comparison to the
  results of the ALICE Collaboration for all charged
  particles~\cite{Abelev:2012hxa, Abelev:2013ala} (solid line) as well
  as for charged pions~\cite{Abelev:2014laa} (dashed line).}
\label{fig:pt}
\end{figure*}

\begin{figure*}[th]
\centering
\includegraphics[width=0.7\textwidth]{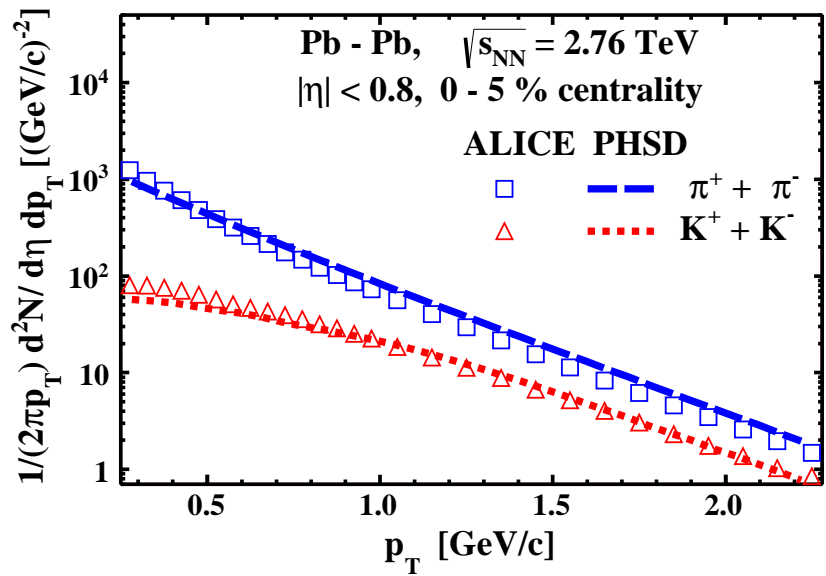}
\caption{Transverse momentum spectra from PHSD for $p_T \leq $ 2 GeV/c
  in comparison to the results of the ALICE Collaboration
  \cite{Abelev:2012hxa, Abelev:2013ala, Abelev:2014laa} for pions and
  kaons.}
\label{fig:pt2}
\end{figure*}

We continue with the transverse momentum spectra for central Pb-Pb
reactions at $\sqrt{s_{NN}}=$~2.76 TeV (0-5\% centrality) which are
compared in Fig.~\ref{fig:pt} with results of the ALICE Collaboration
for all charged particles~\cite{Abelev:2012hxa, Abelev:2013ala} (PHSD:
black solid line) as well as for charged pions~\cite{Abelev:2014laa}
(PHSD: dashed blue line). Note that except for the upgrade in the
PYTHIA version no additional parameters or changes have been
introduced in the PHSD that had been employed before in
Refs.~\cite{CB09, To12, KB12, Kb12b, el-m, photons} from lower SPS up
to top RHIC energies. In this respect the approximate reproduction of
the midrapidity $p_T$ spectra for central collisions over 7 orders of
magnitude in Fig.~\ref{fig:pt} is quite remarkable. A closer look at
the low momentum spectra is offered in Fig.~\ref{fig:pt2} where the
PHSD spectra for pions and kaons are compared to results of the ALICE
Collaboration~\cite{Abelev:2012hxa, Abelev:2013ala,
  Abelev:2014laa}. We recall, furthermore, that the charged particle
pseudo-rapidity density $d N_{ch}/d\eta$ at midrapidity from PHSD
matches well the experimental centrality dependence from ALICE when
displayed as a function of the number of participants $N_{part}$
(cf.\ Fig.~1 in Ref.~\cite{LiLHC}).

\subsection{Differential flow results for Pb-Pb}
\label{sec:flow}

\begin{figure*}[th]
\centering
\includegraphics[width=0.7\textwidth]{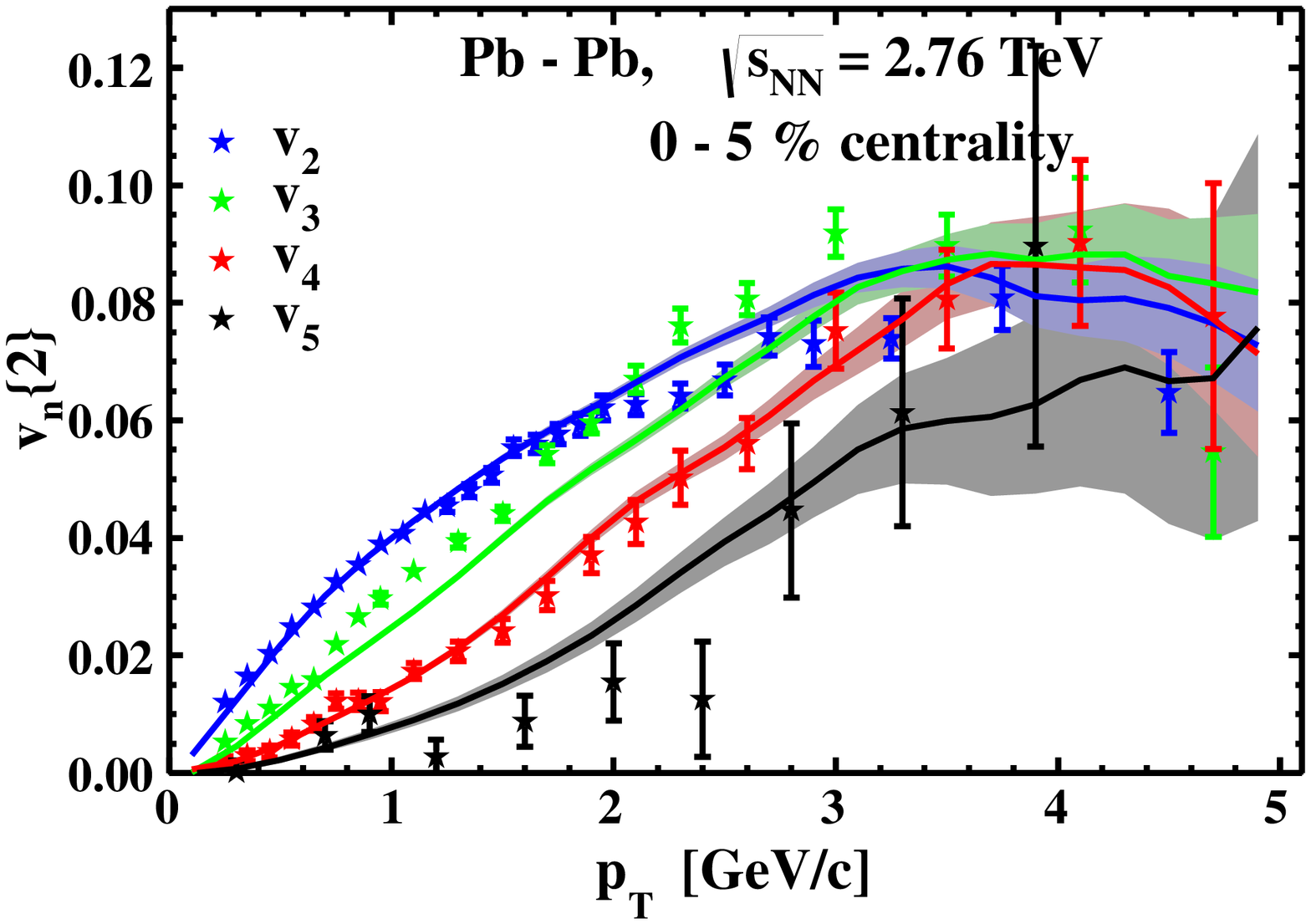}
\includegraphics[width=0.7\textwidth]{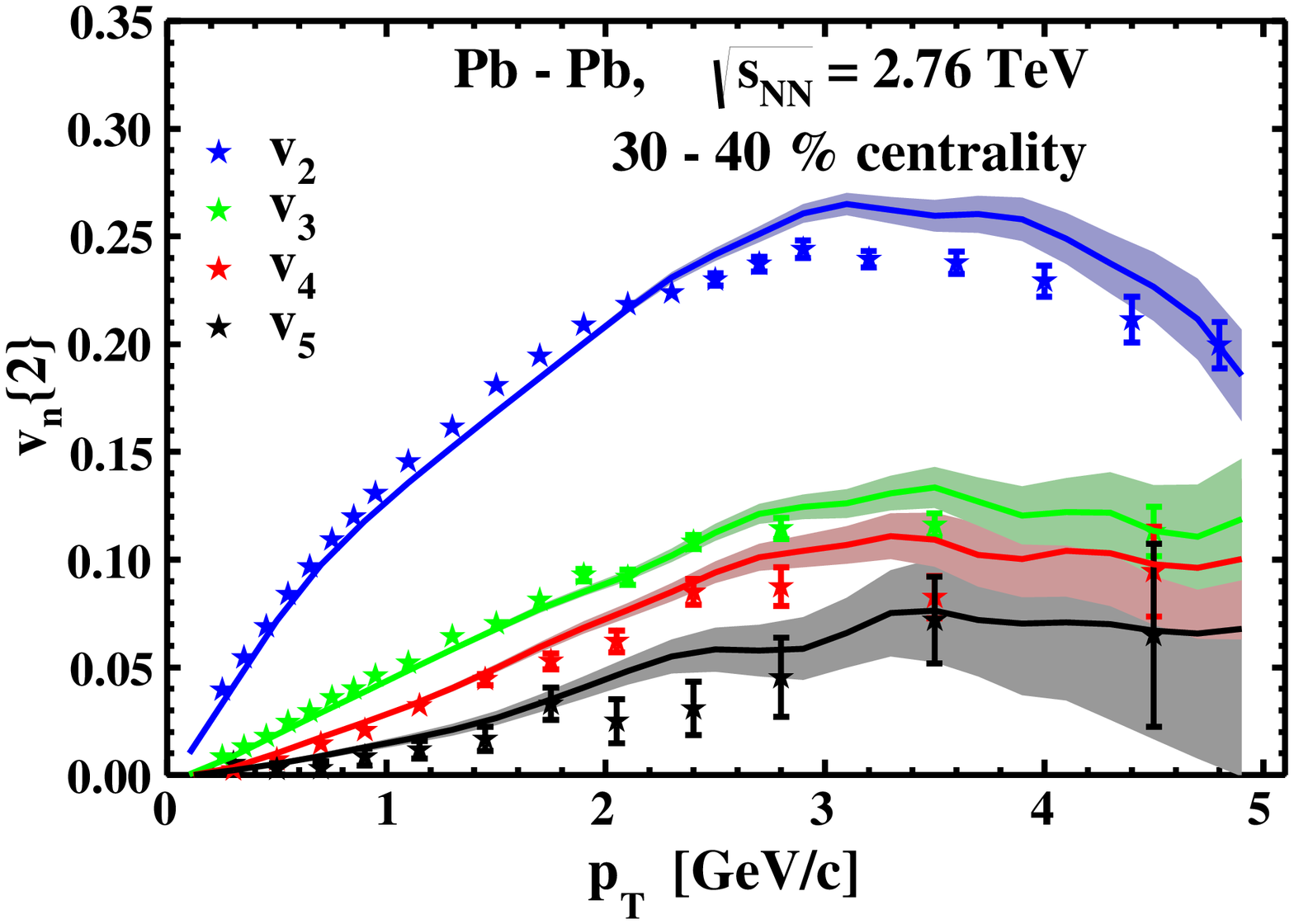}
\caption{The flow coefficients $v_2$, $v_3$, $v_4$ and $v_5$ of all
  charged particles as a function of $p_T$ for the centralities 0-5\%
  ({\it upper part}) and 30-40\% ({\it lower part}). The ALICE data
  have been taken from Ref.~\cite{ALICE:2011ab}. Note the different
  scales for the $v_n$-axis on the upper and lower plots!}
\label{fig:vn}
\end{figure*}

Whereas the transverse charged single-particle spectra compare quite
well with the experimental observation the question remains for the
collective behavior of the system. In this respect the flow
coefficients $v_2$, $v_3$, $v_4$ and $v_5$ of all charged particles
from PHSD are shown in Fig.~\ref{fig:vn} as a function of $p_T$ for
the centralities 0-5\% ({\it upper part}) and 30-40\% ({\it lower
  part}) in comparison to the ALICE data from
Ref.~\cite{ALICE:2011ab}. Anisotropic flow coefficients in both cases
-- i.e.\ experimental data and PHSD calculations -- were obtained from
the two-particle cumulant method~\cite{Voloshin2011} in the central
pseudorapidity window $|\eta|<0.8$ and denoted as $v_n\{2\}$. The PHSD
results for $v_2(p_T), v_3(p_T)$ and $v_4(p_T)$ compare reasonably up
to about 3.5 GeV/c whereas at higher transverse momenta the statistics
is insufficient to draw robust conclusions. This also holds for the
flow coefficient $v_5$ which still is in line with the data within
error bars. It is quite remarkable that the collective behavior is
reproduced not only for semi-central collisions (lower part) but also
for 0-5\% central collisions (upper part) that are more sensitive to
the initial fluctuations~\cite{hyb2}.

These tests indicate that the 'soft' physics at LHC in central A-A
reactions is very similar to the top RHIC energy regime although the
invariant energy is higher by more than an order of
magnitude. Furthermore, the PHSD approach seems to work from lower SPS
energies up to LHC energies for p-p, p-A as well as A-A collisions,
i.e. over a range of more than two orders in $\sqrt{s_{NN}}$.

\section{Probing the initial size of spatial fluctuations with PHSD}
\label{sec:granul}

Having established that the PHSD approach gives results for single
particle spectra as well as collective flow coefficients roughly in
line with experimental observation at LHC energies we now may come to
the central question of this study and address the impact of the
initial size of spatial fluctuations on the collective flow in central
Pb-Pb collisions. As discussed in Ref.~\cite{Kb12b} especially the
triangular flow $v_3$ is sensitive to the initial fluctuations in the
energy density since the average over an ensemble of events at the
same centrality is shape symmetric with vanishing $v_3$. Earlier
studies e.g. in Refs.~\cite{hyb2, hyb5} have shown that the elliptic
flow $v_2$ in semi-peripheral reactions is dominated by the geometry
and less by the initial state fluctuations, however, central
collisions do show a sensitivity to these fluctuations. The same
arguments hold for $v_4$ which is roughly $\sim v_2^2$ (cf.\ Fig.~10
of Ref.~\cite{Kb12b}).

\begin{figure*}[th]
\centering
\includegraphics[width=0.49\textwidth]{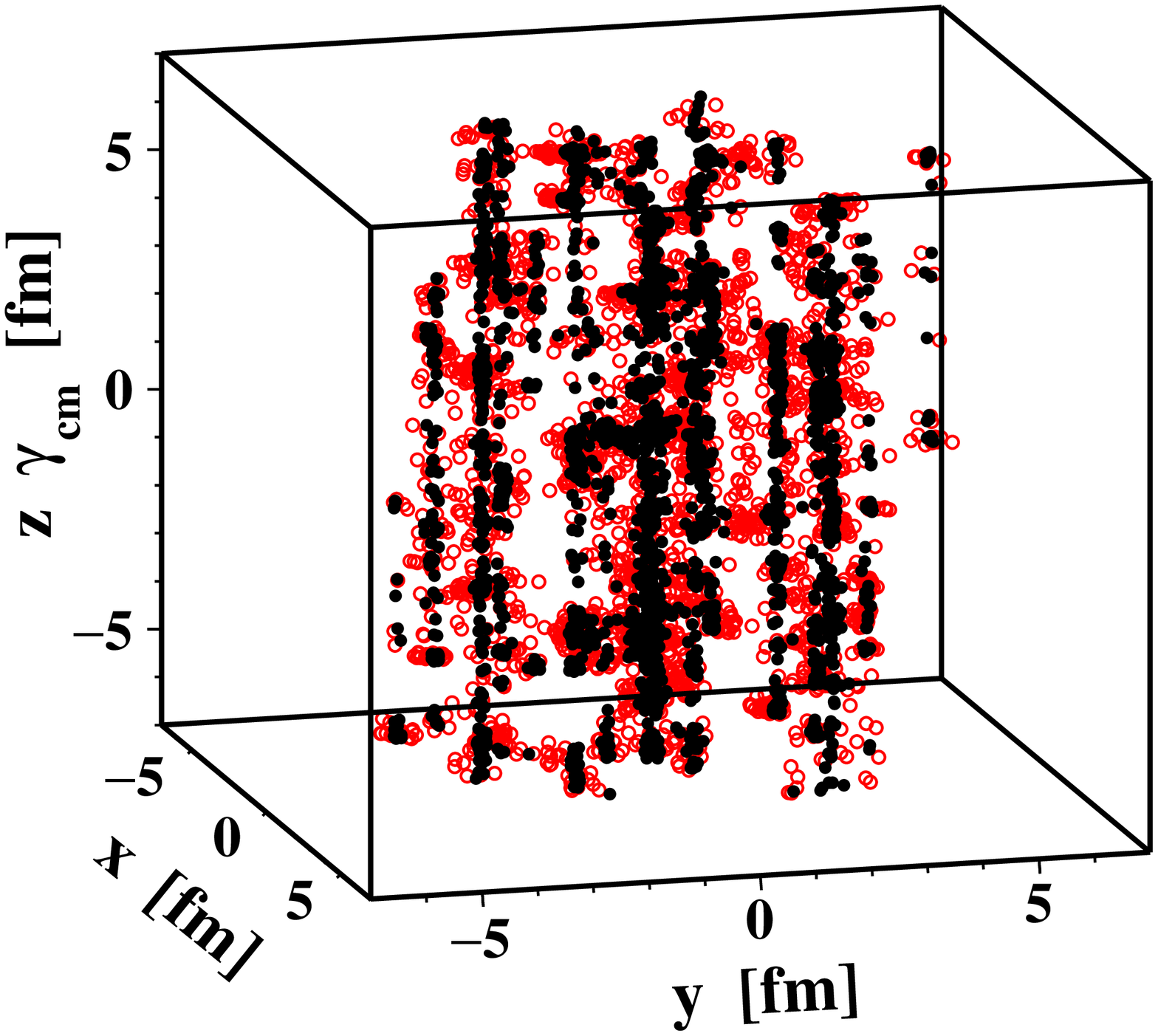}
\includegraphics[width=0.49\textwidth]{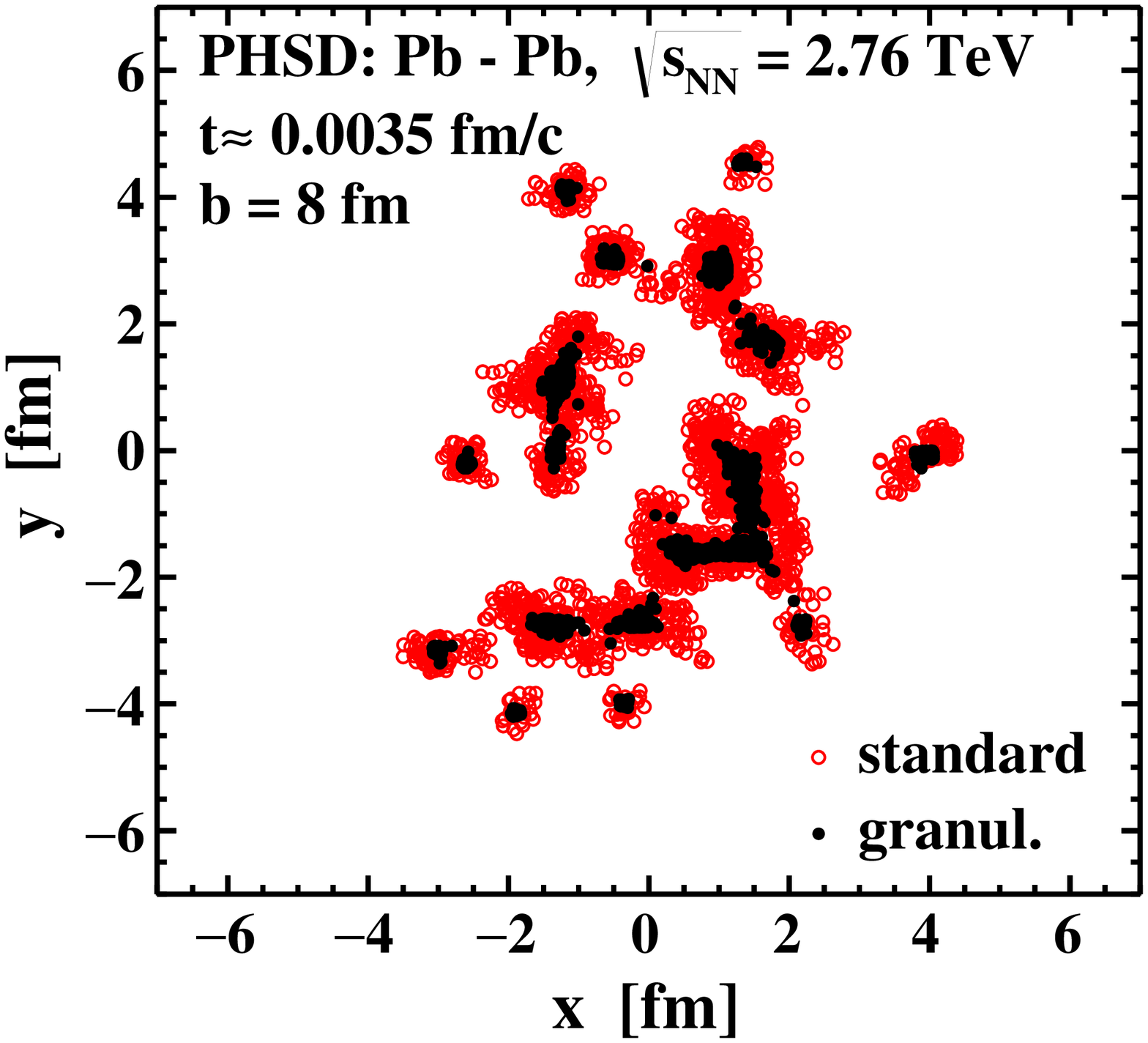}
\caption{Position of the partons in (x,y,z) coordinates ({\it left})
  and their projection on the transverse (x,y) plane ({\it right}) in
  a single Pb-Pb collision at $\sqrt{s_{NN}}=$ 2.76 TeV and impact
  parameter $b$ = 8 fm after full overlap of the nuclei which
  correspond to a time $t=3.5 \times 10^{-3}$ fm/c after the
  contact. Note that the scale in $z$ is multiplied by $\gamma_{cm}
  \approx 1400$. The red open circles correspond to the standard
  scenario of Monte-Carlo Glauber distribution of the initial nucleons
  while the black points show the same partons but shifted to the
  centers of their clusters in order to increase the size of spatial
  fluctuations of the system. This method leaves the shape of the
  event almost unchanged. See text for the details.}
\label{fig:gran}
\end{figure*}

In order to illustrate the local fluctuations in the density we show
in Fig.~\ref{fig:gran} (r.h.s.) the transverse 'particle' density at a
time $t=3.5 \times 10^{-3}$ fm/c after contact of the nuclei for a
Pb-Pb collision at $\sqrt{s_{NN}}=$ 2.76 TeV and impact parameter $b$
= 8 fm in the default PHSD approach (open red circles) which is
comparable to a Monte Carlo Glauber distribution. At this time the two
Pb-nuclei have almost passed through each other, however, the initial
kinetic energy in violent nucleon-nucleon interactions -- as described
by PYTHIA 6.4 -- is converted to a large extent to new
degrees-of-freedom (denoted shortly as partons). This transverse
'lumpy parton' distribution shows separable and overlapping clusters
which in beam (z-)direction have the shape of string-like
configurations (see l.h.s.\ of Fig.~\ref{fig:gran}). Very similar
'lumpy' initial conditions have been presented in
Refs.~\cite{Schenke12,Schenke13,Werner0} and propagated in time by
ideal or viscous hydrodynamics. In PHSD this 'parton' distribution is
stored on a grid with cell-size $\Delta x = \Delta y, \Delta z =
\Delta x /\gamma_{cm}$ where optionally the cell width $\Delta x$ is
chosen between 0.2 fm and 1 fm (depending on the question of
interest). For Pb+Pb collisions the default resolution is $\Delta_x$ =
1 fm which also determines the gradients of the partonic scalar fields
that enter the equations of motion. For our present study we used
$\Delta_x$ = 0.3 fm. In PHSD the actual distributions are point-like
and smeared by a Gaussian of width $\sigma$ from 0.2 fm to 1 fm in
order to achieve particle/energy distributions of different (lower)
size of spatial fluctuations.  However, in order to simulate a
distribution of even higher size of spatial fluctuations -- as
expected for glasma initial conditions~\cite{Schenke12, Schenke13} --
a different strategy has been adopted: a) by means of widely used
cluster algorithms, in particular in statistical physics~\cite{Lu06}
we can identify 'clusters' of particles, evaluate the total 'cluster'
energy and the center of the 'cluster' position; b) in a second more
phenomenological scenario the transversal vectors of all 'particles'
in the cluster relative to the center of the cluster are multiplied by
a common factor $d < 1$ which keeps the 'cluster' position unchanged,
however, increases the local energy density when evaluated on a fine
grid. The full black points in Fig.~\ref{fig:gran} give an example of
the algorithm when increasing the size of spatial fluctuations by a
factor of about three. We mention that our procedure keeps the total
energy and momentum unmodified since only spatial shifts of the
'partons' are involved. However, the shape of each event is
approximately kept such that the harmonics in coordinate space
$\epsilon_n, n=2,3,4,5$ are roughly the same.

Some note of caution has to be added here with respect to the
interpretation of 'particles'. Due to the Heisenberg uncertainty
relation the energy density at $t \approx 3.5 \times 10^{-3}$ fm/c
cannot be specified as being due to 'particles' since the latter may
form only much later on a timescale of their inverse transverse mass
(in their rest frame). More specifically, a jet at midrapidity with
transverse momentum $p_T =$ 100 GeV/c is expected to appear at $t
\approx 2\cdot 10^{-3}$ fm/c while a soft parton with transverse
momentum $p_T \sim 0.5$ GeV should be formed after $t \approx$ 0.4
fm/c. At this time, however, the energy density $\epsilon$ is lower
by more than a factor of 100 due to the dominant longitudinal
expansion. The question, if these degrees-of-freedom in the early time
are coherent gluon fields (glasma), perturbative gluons or virtual $q
\bar{q}$ pairs is presently open. In Ref.~\cite{Konchakovski:2014wqa}
it has been argued that a color glass condensate might be identified
by the rapidity dependence of the $p_T$ of charged particles in p-Pb
reactions while in Ref.~\cite{new} the splitting of the directed flow
for hadrons of opposite charge in mid-central and peripheral
collisions of asymmetric systems (e.g.\ Cu-Au) has been advocated as a
signal for the early presence of electric charges (i.e.\ quarks and
antiquarks).

\begin{figure*}[th]
\centering
\includegraphics[width=0.7\textwidth]{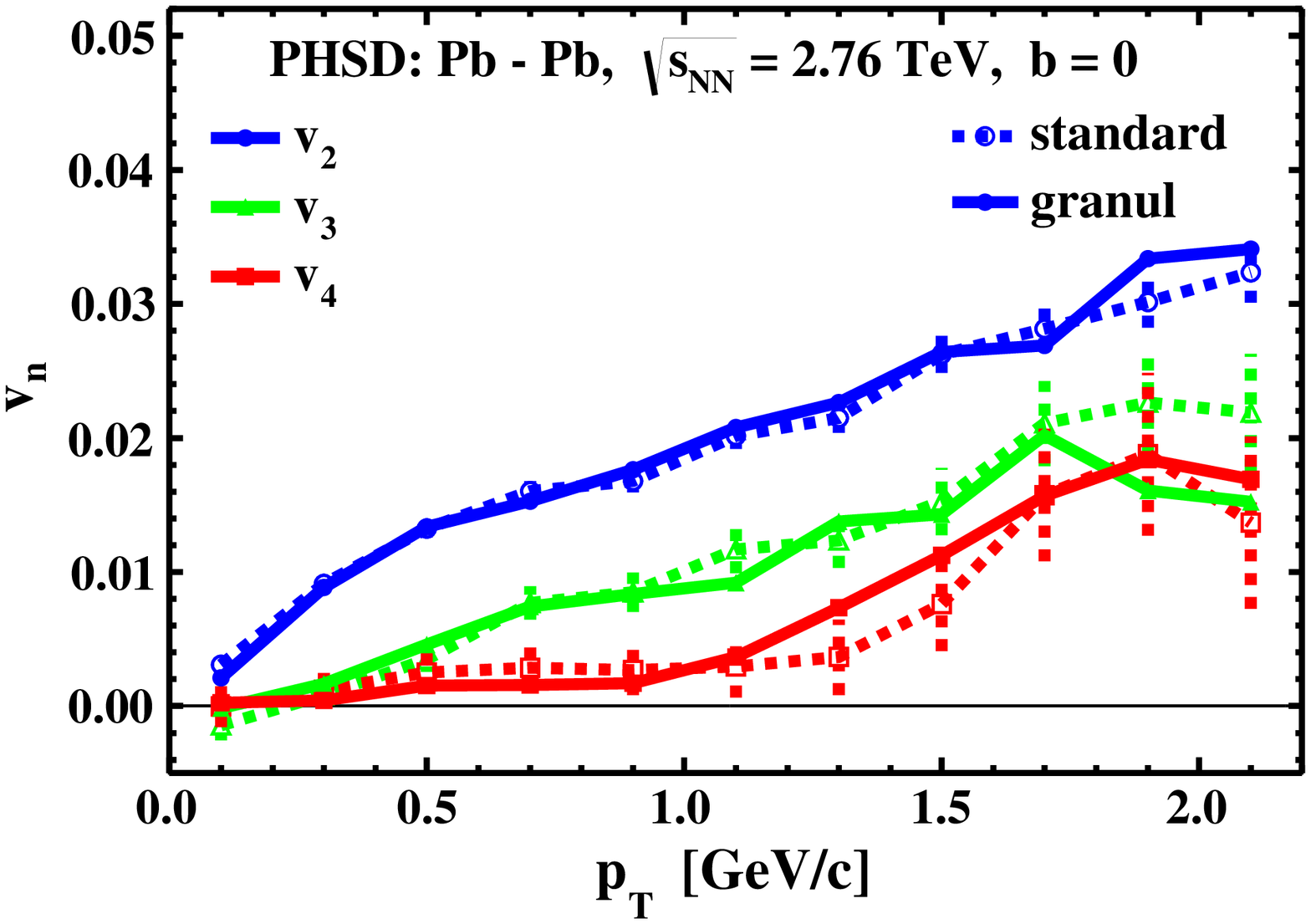}
\includegraphics[width=0.7\textwidth]{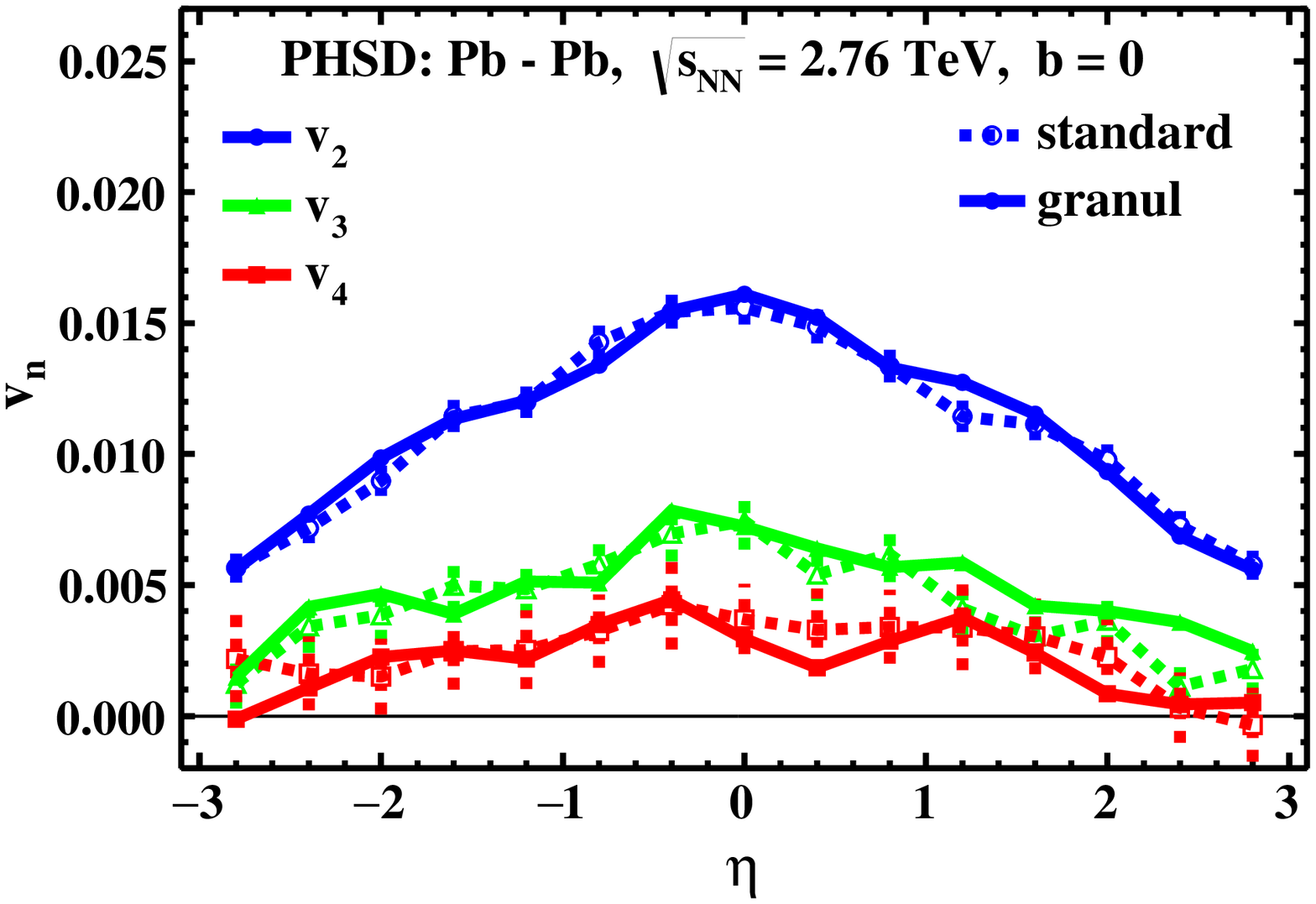}
\caption{Flow coefficients $v_2$, $v_3$ and $v_4$ as a function of
  $p_T$ ({\it upper part}) and $\eta$ ({\it lower part}) for b=0
  collisions of Pb-Pb at $\sqrt{s_{NN}}$ = 2.76 TeV. The results for
  increased size of spatial fluctuations (solid lines) are compared
  with results from the standard PHSD (MC Glauber) initial
  distributions (dotted lines). The error bars denote the statistical
  fluctuations for the default calculation which are of similar size
  for the case with increased size of spatial fluctuations.}
\label{fig:gran_vn}
\end{figure*}

In order to exclude any effect from geometry on the flow coefficients
we now consider very central Pb-Pb collisions with impact parameter
$b$ = 0 at $\sqrt{s_{NN}}=$ 2.76 TeV and perform an event-by-event
analysis. We study two cases: i) standard PHSD evolution with MC
Glauber initial condition which leads to parton distribution as shown
in Fig.~\ref{fig:gran} with red empty dots; ii) we artificially
increase the size of spatial fluctuations of the parton distribution
at $t \approx 3.5 \times 10^{-3}$ fm/c by first identifying the
'cluster' and then specifying the transverse position vector of the
partons relative to the center of the 'cluster'. Then the length of
all parton position vectors is decreased by a factor of three
(scenario b). In this way the size of the 'cluster' decreases
accordingly. The resulting spatial distribution in the transverse
plane is shown by the full black dots in Fig.\ref{fig:gran} (r.h.s.).
The results for the flow coefficients $v_2, v_3$ and $v_4$ are shown
in Fig.~\ref{fig:gran_vn} as a function of transverse momentum $p_T$
(upper part) and pseudorapidity $\eta$ (lower part). Here the default
PHSD calculations are displayed by the dotted lines while the solid
lines result from the same number of events with higher size of
spatial fluctuations. Within the statistical accuracy achieved we find
no apparent sensitivity to the initial size of spatial
fluctuations. Our calculations are in line with the viscous hydro
results from Gale et al.~\cite{Schenke0} for $v_2$ and $v_3$ while
higher moments $v_n$ show are more pronounced sensitivity in the hydro
calculations which, however, are out of reach for the microscopic PHSD
studies. We add in passing that our results for the flow coefficient
$v_5$ in the present case ($b$ = 0) are very 'noisy' in both cases and
do not provide additional information.

\begin{figure*}[th]
\centering
\includegraphics[width=0.7\textwidth]{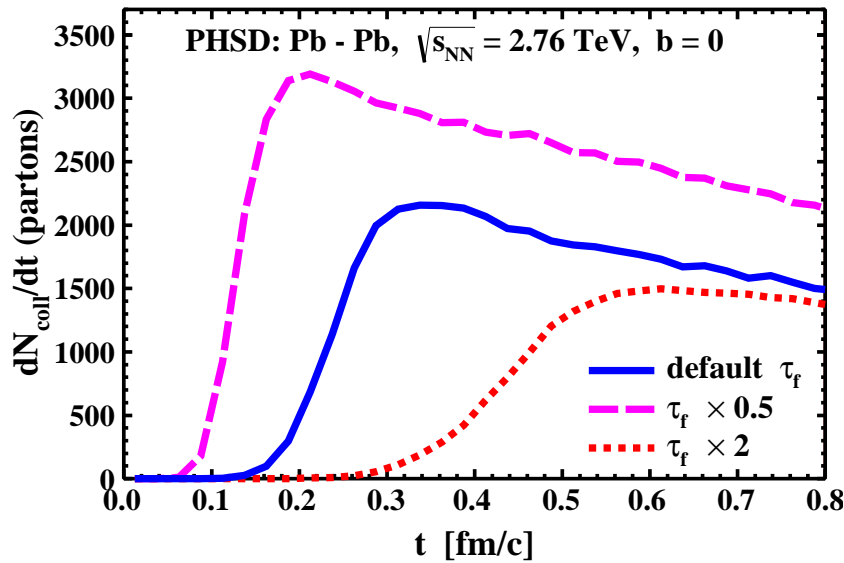}
\caption{Collisional rate of parton-parton interactions from PHSD for
  $b$=0 collisions of Pb-Pb at $\sqrt{s_{NN}}$ = 2.76 TeV for the
  default parton formation time (solid line). The dashed line shows
  the collision rate when artificially decreasing the formation time
  by a factor of two while the dotted line is obtained for a twice
  larger formation time.}
\label{fig:rate}
\end{figure*}

The reason for the insensitivity to the initial size of spatial
fluctuations is partly due to fact that the system in the very early
stage is almost collision-less since only formed partons -- with a
'dressed' propagator -- interact in PHSD where the formation time of a
parton (in its rest frame) is given by $\tau_f = 1/M_T$ where $M_T$
denotes the transverse mass of the parton.  This is demonstrated
explicitly for a central Pb-Pb collision at $\sqrt{s_{NN}}$ = 2.76 TeV
for $b$ = 0 fm in Fig.~\ref{fig:rate} (solid line) where the partonic
interaction rate from PHSD is displayed as a function of time $t$ from
contact. It is clearly seen that the interaction rate is very low for
about 0.2 fm/c; during this time the local clusters have increased in
transverse diameter by about 0.6 fm such that the energy distributions
between the default and squeezed initial conditions become
similar. Furthermore, as mentioned before, the initial event shape in
coordinate space is not changed very much when increasing the size of
spatial fluctuations. In ideal hydrodynamics one thus would expect
very similar flow coefficients $v_n$ ($n$=2,3,4).

\begin{figure*}[th]
\centering
\includegraphics[width=0.7\textwidth]{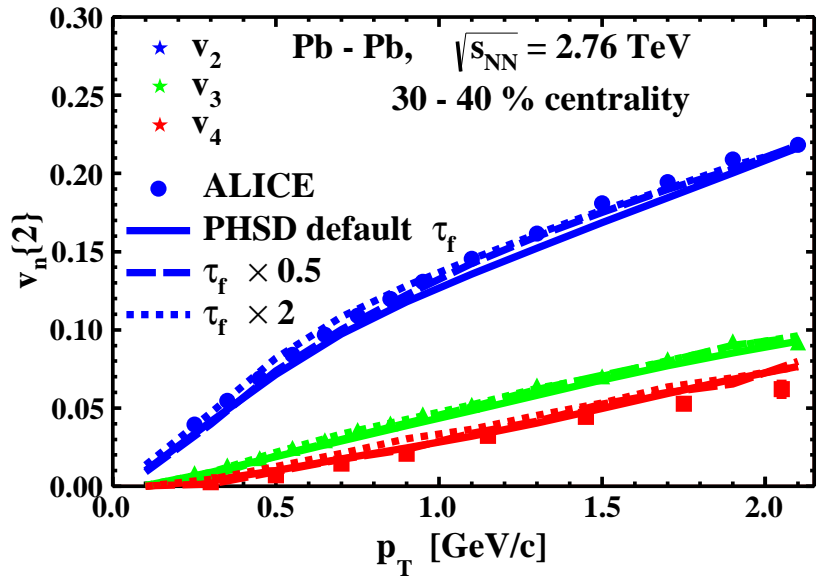}
\caption{Flow coefficients $v_2$, $v_3$ and $v_4$ as a function of
  $p_T$ for 30-40\% central Pb-Pb collisions at $\sqrt{s_{NN}}$ = 2.76
  TeV for the default formation time (solid line) as well as twice
  smaller (dashed) and twice larger ones (dotted).}
\label{fig:vn_ftime}
\end{figure*}

In order to explore the effect of different parton formation times we
artificially decrease (or increase) $\tau_f$ by a factor of two and
accordingly find the parton collision rate to start earlier (later) as
shown in Fig.~\ref{fig:rate} by the dashed (dotted) line.  However for
all the cases the flow coefficients $v_2$, $v_3$ and $v_4$ do not
change within error bars up to $p_T\approx$ 2 GeV/c as seen in
Fig.~\ref{fig:vn_ftime} for 30-40\% central Pb-Pb collisions at
$\sqrt{s_{NN}}$ = 2.76 TeV. This suggests that the similarity of the
global event shape (within enhanced size of spatial fluctuations)
dominantly drives the flow coefficients.

\section{Conclusions}
\label{sec:conclusions}

In this study the parton-hadron-string dynamics (PHSD) approach has
been employed in the LHC energy range for Pb-Pb collisions as well as
p-Pb collisions at $\sqrt{s_{NN}}$ = 5.02 TeV. We find that this
approach works also reasonably for Pb-Pb collisions at $\sqrt{s_{NN}}$
= 2.76 TeV with respect to charged particle spectra as well as
collective flow coefficients $v_2, v_3$, $v_4$ and $v_5$ for different
centralities with a quality comparable to that achieved at RHIC
energies before~\cite{CB09, To12, KB12, Kb12b, el-m, photons}. Our
finding implies that the 'soft' physics in Pb-Pb collisions at the LHC
and Au-Au interactions at the top RHIC energies -- despite a factor of
$\sim$ 14 in $\sqrt{s_{NN}} $ -- is very similar and in line with the
dynamical quasiparticle model (DQPM) that defines the parton
properties for PHSD in equilibrium. This finding is common with
earlier studies using viscous hydro approaches with varying initial
conditions~\cite{Heinz}.

The particular question addressed in this study has been the
dependence of the collective flow observables to the initial size of
spatial fluctuations in the parton density or energy density. The PHSD
calculations have shown no sensitivity to the initial size of spatial
fluctuations for the flow harmonics $v_2$ to $v_4$ which is due to the
low interaction rate in the initial nonequilibrium stage in PHSD
($\sim$ 0.3 fm/c) where effects from different sizes of spatial
fluctuations are already washed out to some extent. On the other hand
our method for changing the size of spatial fluctuations keeps the
event shape in coordinate space approximately invariant
(cf.\ Fig.~\ref{fig:gran}) which - in line with hydrodynamics - leads
to very similar flow coefficients $v_n$ in momentum space. We mention
that the low interaction rate in this very early phase in PHSD is
common with the CGC concept and thus does not allow to disentangle or
determine the effective degrees-of-freedom in this 'pre-hydro' phase.

\section*{Acknowledgments}
The authors are thankful to E.~L.~Bratkovskaya and O.~Linnyk for
illuminating discussions. This work in part was supported by the LOEWE
center HIC for FAIR as well as by BMBF. V.~D.\ was also partly
supported by the Heisenberg-Landau grant of JINR and the Russian
Ministry of Science and Education, research project RFMEFI61614X0023.


\end{document}